\documentclass[aps,amssymb,preprint,a4paper,longbibliography]{revtex4}
\usepackage[T1]{fontenc}
\usepackage[english]{babel}
\usepackage{braket}
\usepackage{epsfig}
\usepackage{graphicx}
\usepackage{amsmath,amsfonts,amssymb}
\usepackage{dsfont}
\usepackage{booktabs}
\usepackage{units}
\usepackage{natbib}
\usepackage{xcolor}
\usepackage{multirow}
\usepackage{tikz,pgfplots}

\usetikzlibrary{patterns,shadows,trees,calc}
\usepgfplotslibrary{units}

\pgfplotsset{compat=1.8}

\begin{document}

\newcommand{\wignerj}[6]{\mbox{$\left( \begin{array}{ccc} #1 & #2 & #3 \\ #4 & #5 & #6 \end{array} \right)$}}
\newcommand{\redume}[3]{\mbox{$( #1 || #2 || #3 )$}}

\setlength{\tabcolsep}{12pt}

\definecolor{diplom1}{rgb}{0.0 0.4 1.0}
\definecolor{diplom2}{rgb}{0.0 0.0 0.6}
\definecolor{diplom3}{RGB}{153,0,0} 

\title{Effect of Spin-Orbit Coupling on Decay Widths of Electronic Decay Processes}

\author{Elke Fasshauer$^{\text{a}}$}
\email[]{Email:Elke.Fasshauer@gmail.com}

\affiliation{$^{\text{a}}$Department of Physics and Astronomy,
Aarhus University, Ny Munkegade 120, 8000 Aarhus, Denmark}

\date{\today}

\begin{abstract}
Auger-Meitner processes are electronic decay processes of energetically low-lying
vacancies. In these processes, the vacancy is filled by an electron of
an energetically higher lying orbital, while another electron is simulataneously
emitted to the continuum.
In low-lying orbitals relativistic effects can not even be neglected for light
elements. At the same time lifetime calculations are computationally expensive.
In this context, we investigate which effect spin-orbit coupling has on
Auger-Meitner
decay widths and aim for a rule of thumb for the relative decay widths of
initial states split by spin-orbit coupling.
We base this rule of thumb on Auger-Meitner decay widths
of Sr$4p^{-1}$ and Ra$6p^{-1}$
obtained by relativistic FanoADC-Stieltjes calculations.
\end{abstract}

\maketitle


\section{Introduction}
Electronic decay processes are initiated by a sub-outer valence ionization
or excitation. The system relaxes by filling the vacancy with an electron and
transferring the excess energy to another electron of the system, which is emitted.
One examples of electronic decay processes is the
well known Auger process. It was first discovered
by Lise Auger-Meitner \cite{Meitner22} and later rediscovered
by Pierre Auger \cite{Auger23}. In order to give credit where credit is due
{and following the renaming suggestion of Ref. \cite{renameAuger19}
we will refer to it as the Auger-Meitner process.}
It is found in a variety of different systems like atoms, molecules and solids.
As a consequence, Auger-Meitner spectroscopy
is used for surface analysis in metallurgy,
quality analysis of microelectronics
as well as for basic studies of chemical reaction mechanisms \cite{AES_Seah_86}.
Since the advent of short XUV pulses, it has been a test process for
time-resolved measurements
{\cite{Drescher02,Smirnova03}}
and is often observed as a side product of modern x-ray spectroscopies
\cite{Greczynski20}.
Another relevant and interesting process is the family of Interparticle
Coulombic Decay processes (ICD) \cite{Cederbaum97,Hergenhahn11,Jahnke15}.
A deeper understanding of the effects that determine the decay widths
of electronic decay processes, that
can lead to rule of thumbs, which do not require expensive lifetime
calculations will
be helpful for the interpretation of the observed spectra.
This article aims to provide such a rule of thumb for relative
{Auger-Meitner}
decay widths
of initial states split by spin-orbit coupling.
In the following, we will assume processes initiated by ionization.

In order to occur, two criteria need to be fulfilled: the energy and the coupling
criterion.
To fulfill the energy criterion the final state energy is required
to be lower than
the energy of the singly ionized initial state. If this is not the case, the
channel defined by a certain doubly ionized final state
is closed and the corresponding fragments of the
channel are not observed after the decay.
To fulfill the coupling criterion, the decay process needs to be fast enough
to prevail over other energetically accessible decay pathways like radiative
relaxation or coupling to nuclear degrees of freedom.
It hence contains
the information whether an energetically allowed process can be expected
to be observed experimentally or not.
Therefore, a typical study of electronic decay processes consists of two parts:
\begin{itemize}
 \item determination of the kinetic energy of the secondary electron
       and, as a consequence, which decay channels are open
 \item calculation of the decay width $\Gamma=\frac{\hbar}{\tau}$, which
       is proportional to the decay rate $\frac{1}{\tau}$ and
       inversely proportional to the lifetime $\tau$
\end{itemize}

The primary ionization often removes an electron from an atomic core, where
relativistic effects are stronger than for valence electrons. The relativistic
effects can therefore be expected to play a crucial role for the understanding
of the systems' lifetimes.
Phenomenologically, the relativistic effects can be divided  
into spin-orbit coupling and scalar-relativistic effects. The spin-orbit       
coupling requires the system to be described in terms of the total angular     
momentum $j$ rather than the orbital momentum $l$ and the spin momentum $s$.   
Thereby the non-relativistically degenerate states of one particular $l$       
value are split into two states with $j=l\pm s$ of different energies          
\cite{ReiherWolf09}.                                            
The scalar-relativistic effects result in spatial contractions of all orbitals on
one-electron systems. In many-electron systems, however, those orbitals with density close
to the nucleus are more strongly contracted than others. They thereby
shield the positive charge of the nucleus from the electrons in other orbitals,
which are therefore effectively spatially decontrated compared to the non-relativistic
solutions. Therefore, as a rule of thumb, $s$ and $p$ are spatially contracted while
$d$ and $f$ orbitals are spatially decontrated. \cite{ReiherWolf09}.

The initial and final state energies can be obtained using      
a variety of quantum chemical approaches.
{
One is the literature as the       
Algebraic Diagrammatic Construction \cite{Schirmer_book,Schirmer82_1,Schirmer83,Schirmer91,Schirmer98, 
Mertins96_1}} (ADC), which is also                               
available for a fully relativistic treatment                    
\cite{Pernpointner04_1,Pernpointner04_2,Pernpointner10_1}.
The calculation of the respective lifetimes is a challenging task, because it
requires the description of both bound and continuum electrons. The bound electrons
are best described by wavefunctions with $\mathcal{L}^2$ boundary conditions,
while continuum electrons far away from the atom are best described by
plane waves. This imposes the technical choice between a) describing the entire
atom using an $\mathcal{L}^2$ basis, b) describing the entire process using
a grid, or c) describing the bound electrons using an $\mathcal{L}^2$ basis
and the continuum electron using a grid.
Either     
of these approaches faces difficulties in either describing the bound or the  
continuum states or some artificially constructed interface region.
Traditionally, most quantum
chemical program packages are based on $\mathcal{L}^2$ bases and we therefore
choose to describe the electronic decay processes using a large $\mathcal{L}^2$
basis for convenience.

The quantum chemical methods, which have been developed for a relativistic
description of the decay widths are the Multichannel Multi-Configurational
Dirac-Fock (MMCDF) \cite{Fritzsche11} and the FanoADC-Stieltjes
{\cite{Averbukh05,Fasshauer15_1}}.
The MMCDF is limited to atomic systems and strongly depends on
the manual selection of CI (Configuration
Interaction) components to be included in the description of initial and      
final states.
The FanoADC-Stieltjes is based on the ADC and not limited to spherically symmetric
systems.
It is furthermore size-consistent and includes terms up to third order
in perturbation theory yielding the interaction between initial and final state
without actually calculating the states themselves.
It is therefore a good compromise between
accuracy and computational cost. At the same time, the configurations needed for the
accurate description of initial and final states are determined automatically up to
the implemented order of the perturbed wavefunction.
The FanoADC-Stieltjes approach is implemented in the relativistic
quantum chemistry package
DIRAC \cite{DIRAC17}, which allows to use different Hamiltonians while keeping
all other parameters constant. This allows us to computationally only consider
scalarrelativistic effects using
{Dyall's spinfree Hamiltonian \cite{Dyall94}}
as well as
performing a fully relativistic calculation using the four-component
Dirac-Coulomb (DC) Hamiltonian taking both scalarrelativistic and
spin-orbit coupling into account.
By comparing the results, we can extract the effect caused by spin-orbit coupling.
We therefore choose the FanoADC-Stieltjes approach for our calculations.

In this work we focus on the atomic Auger-Meitner process     
in order                                      
to exploit basic knowledge about the influence of relativistic effects,
but we expect the conclusions to hold for molecular Auger-Meitner process and
Interparticle Coulombic Decay (ICD) processes
\cite{Cederbaum97,Marburger03,Hergenhahn11,Jahnke15} as well.
The Auger-Meitner decay process initiated by a photoionization
can most generally be described by:
\begin{equation*}                                               
 A \quad \xrightarrow{h\nu}\quad (A^+)^* + e^-_\text{ph} \quad       
    \xrightarrow{\text{Auger-Meitner}} \quad A^{2+} + e^-_\text{ph} + e^-_\text{sec}     
\end{equation*}                                                 
                                                                
A system $A$ is ionized, while the photo-electron $e^-_\text{ph}$ is emitted.      
Afterwards, the actual Auger-Meitner
process of the initial state $A^+$ can occur.    
An electron from an outer                                       
shell fills the vacancy and the excess energy is instantaneously transferred  
to another (secondary) electron $e^-_\text{sec}$, which
is subsequently emitted. The final state of the decay process   
is to be described by a doubly charged atom $A^{2+}$ and the secondary        
electron in the continuum.
{In the special case when the vacancy filling electron or the secondary
electron stem from an orbital
of the same shell as the initial vacancy the process is called Coster-Kronig decay.
When both are from the same shell, the process is referred to as super-Coster-Kronig
decay. These Coster-Kronig decays are characterized by very large decay widths.
\cite{Coster35} We will discuss these special cases separately from the 
Auger Meitner process in this paper.
}

We have previously shown, how
scalar-relativistic effects influence the decay widths of the Auger-Meitner
process in
noble gas atoms. \cite{Fasshauer15_1}
After primary ionization from the $(n-1)d$ orbitals the nobel
gas atoms decay to $np^{-2}$, $np^{-1}ns^{-1}$ and $ns^{-2}$ final states.
We observed that the Auger-Meitner decay widths increased by up to \unit[326]{\%}
for radon
by including scalar-relativistic effects in the calculation. This
dramatic increase could be explained by the larger spatial
overlap of the orbitals
involved in the decay compared to non-relativistic calculations due to
a contraction of the $s$ and $p$ orbitals of the final states.\\
However, the fully relativistic calculation resulted in different decay widths
for the different $d_{3/2}^{-1}$ and $d_{5/2}^{-1}$ initial states split by
spin-orbit coupling (see Fig.~5 of Ref.~\cite{Fasshauer15_1}).
The aim of this work is therefore to investigate the
influence of spin-orbit coupling on the decay widths of electronic decay processes.
For this purpose, we will study the Auger-Meitner processes of earth alkaline atoms
after primary ionization from the $(n-1)p$ orbitals. The earth alkaline elements
have the benefit of a single and closed shell $ns^{-2}$ final state.
This might allow us to purely observe how the spin-orbit splitting of
the different initial states affects the Auger-Meitner decay width.
{
Moreover, in the single particle picture, a Coster-Kronig decay is possible for neither
of the initial states and thus the pure Auger-Meitner process, where the initial
states have an equal number of possible decay channels, can be observed.
}
Having only one final state, which is not affected by spin-orbit coupling,
reduces the complexity of the analysis of the results significantly.

The paper is structured as follows:
in section \ref{section:theory} we recapitulate the basics of the
FanoADC-Stieltjes method. We then give the computational details for our
ab initio calculations in section \ref{section:computational}. We present the
results and their interpretation in section \ref{section:results}
and conclude in section \ref{section:conclusions}.


\section{Theory}
\label{section:theory}

Following Wentzel \cite{Wentzel27} and later Feshbach \cite{Feshbach58,Feshbach62}
and Fano \cite{Fano61}
the decay width of a decay process initiated by a
primary ionization is given by 

\begin{equation} \label{equation:Fano_golden}
  \Gamma = \sum_\beta 2\pi
           \left| \braket{\Phi|\hat{V}|\chi_{\beta,\varepsilon}} \right|^2 .
\end{equation}

Here, $\ket{\Phi}$ and $\ket{\chi_{\beta,\varepsilon}}$ denote the initial and
final state, respectively. $\hat{V}$ is the interaction operator of the
initial and final states, which in Feshbach's definition is known as $H_{PQ}$.
The index $\beta$ refers to the different
decay channels and $\varepsilon$ denotes the energy of the final state.
Eq. (\ref{equation:Fano_golden}) thereby connects the metastable initial
and the continuum final states. They are constructed by partitioning the
Hamiltonian into two subspaces. The initial (final) state is then an
eigenfunction of this initial (final) state sub-space Hamiltonian.
However, finding proper solutions to both the initial and the final
states on an equal footing is a non-trivial task, because they adhere to
different boundary conditions. Since the final state depends on the energy
of the emitted electron, any approach needs to either determine the continuum
state or to mimic the final state using $\mathcal{L}^2$-functions.
While the continuum functions are normalized with respect to their energy

\begin{equation}
 \braket{\chi_\varepsilon| \chi_{\varepsilon'}} = \delta(\varepsilon-\varepsilon')
\end{equation}

the $\mathcal{L}^2$ approach is based on a discrete set of final states
$\ket{\tilde{\chi}_{\tilde{E}}}$
which adhere to different boundary
conditions and are normalized with respect to space (see e.g. \cite{Craigie14})
\begin{equation}
 \braket{ \tilde{\chi}_{\tilde{E}_i} | \tilde{\chi}_{\tilde{E}_j} } = \delta_{ij}.
\end{equation}

Because of this different normalization the decay widths are not amenable to
a direct calculation. As first proposed by Hazi \cite{hazi1978}, for
autoionization processes such difficulties
can be solved by using the
Stieltjes-Chebyshev moment theory also called Stieltjes imaging
\cite{Langhoff76,Corcoran77,MuellerPlathe90}.
It relies on the observation that the moments of order $k$ of the projected final
state Hamiltonian $H_f$

\begin{equation}
 \mu_k = \braket{ \Phi | \hat{V} H^k_f \hat{V} | \Phi }
\end{equation}

calculated from the determined
discrete pseudo-spectrum are good approximations to the moments determined
from the real continuum states.
This can be shown by inserting the resolution of identity for
the continuum states

\begin{equation}
 \mu_k = \sum_i \varepsilon_i^k
         \left| \braket{ \Phi | \hat{V} | \chi_{i,\varepsilon} } \right| ^2
       + \int\limits_{E_{0}}^{\infty} \varepsilon^k
         \left| \braket{\Phi|\hat{V}|\chi_{\varepsilon}} \right|^2 \mathrm{d}\varepsilon  .
\end{equation}

Since the non-zero contribution to the coupling matrix elements in the
Feshbach-Fano approach stems only
from an interaction region of finite size, where the $\mathcal{L}^2$ final
state functions are nonvanishing, we may replace the expansion
$\sum\limits_i \ket{\chi_{i,\varepsilon}} \bra{\chi_{i,\varepsilon}}
 + \int \mathrm{d}\varepsilon \ket{\chi_\varepsilon} \bra{\chi_\varepsilon}$
by its $\mathcal{L}^2$ approximation
$\sum\limits_j \ket{\tilde{\chi}_{\tilde{E}_j}} \bra{\tilde{\chi}_{\tilde{E}_j}}$
(see \cite{Reinhardt79})

\begin{equation}
 \label{eq:moment_discrete}
 \mu_k \approx \sum\limits_j \tilde{E}_j ^k
         \left| \braket{\Phi|\hat{V}|\tilde{\chi}_{\tilde{E}_j}}  \right|^2 .
\end{equation}

Then the decay width can be determined through a series of
consecutive approximations to the moments of increasing order $k$.

To achieve this kind of description, we choose the relativistic
FanoADC-Stieltjes approach.
Here, a proper selection of $2h1p$ intermediate state configurations are used
for the description of the continuum final state,
while the rest is used for the description
of the initial state.
The resulting discrete pseudo-spectrum
is then subject to a Stieltjes imaging procedure.
An exhaustive description of the method can be found in Refs. \cite{Fasshauer15_1}
and \cite{Fasshauer_thesis}.

\section{Computational Details}
\label{section:computational}
The Auger-Meitner
decay widths were calculated with the relativistic FanoADC-Stieltjes
method
implemented in the relativistic quantum chemical program DIRAC \cite{DIRAC17}.
We included up to third order contributions of perturbation theory and additional
constant diagrams.
For each element four-component calculations based on the
Dirac-Coulomb (DC) Hamiltonian
and scalarrelativistic spinfree calculations were
performed for both the $(n-1)p_{1/2}$ and $(n-1)p_{3/2}$ initial states.
Dyall's cv4z basis sets \cite{Dyall4s-7s09} were augmented with additional diffuse
5s5p5d3f
basis functions following the Kaufmann-Baumeister-Jungen approach
\cite{Kaufmann89}.
The resulting moments were checked for numerical instabilities.
Only those moments, without numerical instabilities entered the interpolation
scheme for the determination of the decay widths
{as described in Ref. \cite{Fasshauer15_1}}.\\
The radial orbital densities of the ions were calculated using GRASP
\cite{Dyall89,Parpia96}.

\section{Results}
\label{section:results}

\begin{figure}[h]
 \centering
 \includegraphics[width=\columnwidth]{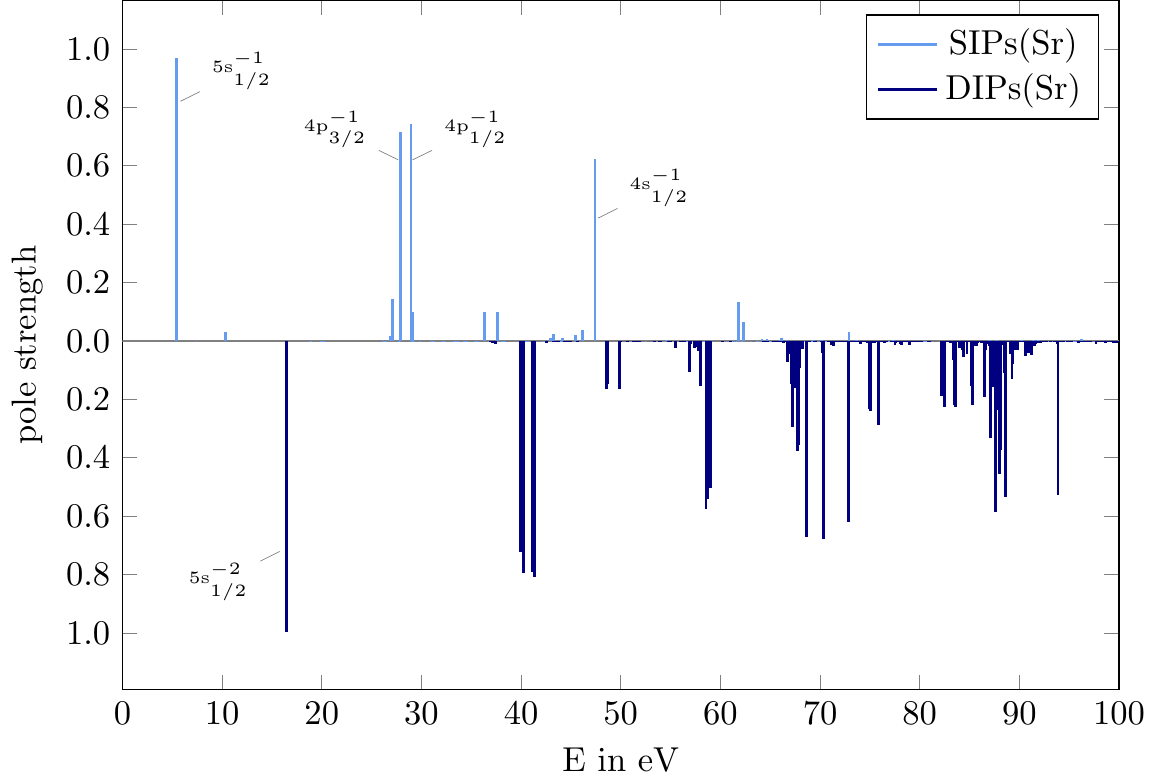}
 \caption{Comparison of the single (SIP) and double (DIP) ionization spectra
          of the strontium obtained by a DC-ADC calculation.}
 \label{fig:sdip}
\end{figure}

In order to analyze the decay processes, we first discuss the energetic
accessibility of different decay channels and then present the corresponding
Auger-Meitner decay widths.
In Fig. \ref{fig:sdip} we present the computed
single and double ionization spectra
of strontium obtained by DC-ADC calculations. The main peaks of the
ionization
from the Sr$4p_{3/2}$ and the Sr$4p_{1/2}$ have single ionization potentials (SIPs)
of \unit[28.277]{eV} and \unit[29.402]{eV}, respectively. The experimental
values of \unit[28.21]{eV} and \unit[29.17]{eV} are very close and
{verify}
the applicability
of the chosen method \cite{Schmitz76}.
They are characterized by pole-strengths, which in this implementation
is defined as
the sum of the absolute squares of the $1h$ coefficients
\cite{Trofimov05}, of 0.76 and 0.80
(see Table \ref{tab:widths}). We will at this point treat them as single
configurations and analyze them in detail later.
These two single ionization energies are higher
than only one double ionization potential (DIP) of \unit[16.430]{eV},
which is to \unit[99.7]{\%} characterized by the double ionization from
the $5s$ valence. The Auger-Meitner process is therefore energetically accessible and
results in a single final state.
For radium the spectra are qualitatively the same and
lead to the same conclusion.

\begin{table}[htb]
 \centering
 \caption{Ionization energies, pole-strengths (ps) and decay widths $\Gamma$ of
          different Auger-Meitner initial states of strontium and radium obtained
          by relativistic FanoADC-Stieltjes calculations.}
 \begin{tabular}{lrrr}
  \toprule
   initial state    & energy $[\unit{eV}]$ & ps & $\Gamma [\unit{meV}]$\\
  \midrule
   Sr spinfree      & 28.599 & 0.78 &   0.56\\  
   Sr$4p_{1/2,1/2}$ & 29.402 & 0.80 &   0.10\\
   Sr$4p_{3/2,1/2}$ & 28.277 & 0.76 &   1.23\\
   Sr$4p_{3/2,3/2}$ & 28.277 & 0.76 &   1.17\\
  \midrule
   Ra spinfree      & 21.836 & 0.49 &  28.56 \\  
   Ra$6p_{1/2,1/2}$ & 25.494 & 0.78 &   0.26\\
   Ra$6p_{3/2,1/2}$ & 19.267 & 0.50 &  93.16 \\
   Ra$6p_{3/2,3/2}$ & 19.267 & 0.50 &  98.86\\
  \bottomrule
 \end{tabular}
 \label{tab:widths}
\end{table}

\begin{figure}[h]
 \centering
 \includegraphics[width=0.95\columnwidth]{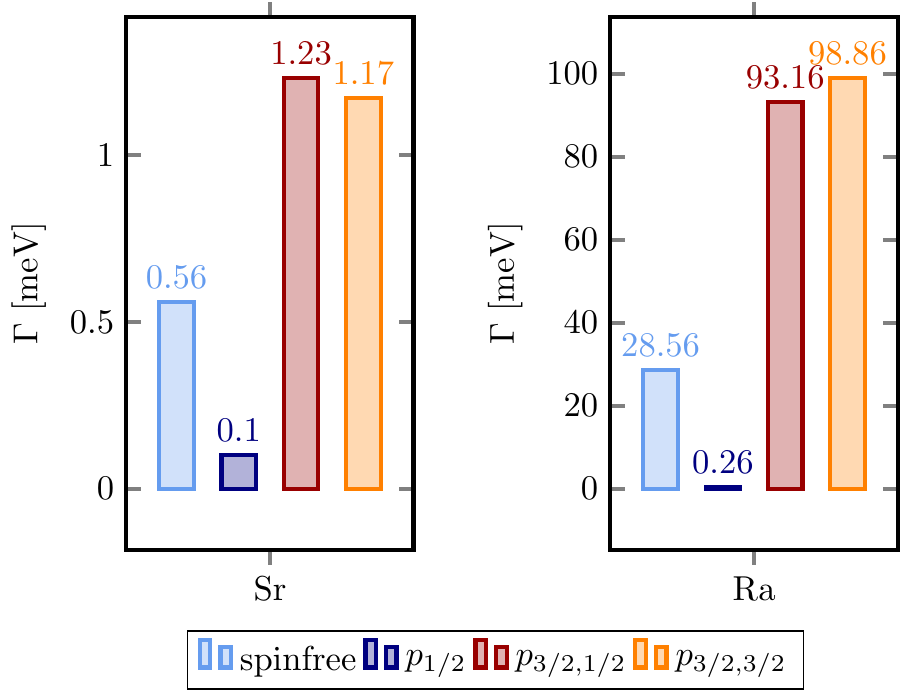}
 \caption{Decay widths of strontium and radium after primary ionization from
          an $(n-1)p$ orbital obtained by relativistic FanoADC-Stieltjes
          calculations with different Hamiltonians. The spinfree Hamiltonian
          includes only scalarrelativistic effects, while the four-component
          DC Hamiltonian also includes spin-orbit coupling, which gives rise to
          three possible initial states: the $p_{1/2}^{-1}$ as well as the
          degenerate $p_{3/2,1/2}$ and $p_{3/2,3/2}$ initial state.
          The decay width of the $p_{1/2}$
          initial state is significantly lower than for either of the
          $p_{3/2}$ initial
          states.}
 \label{fig:gamma}
\end{figure}

We show the corresponding decay
widths of strontium and radium obtained by relativistic FanoADC-Stieltjes
calculations in Table \ref{tab:widths}
for the scalarrelativistic spinfree $(n-1)p^{-1}$ as well as the
fully relativistic $(n-1)p_{1/2,1/2}^{-1}$, $(n-1)p_{3/2,1/2}^{-1}$ and
$(n-1)p_{3/2,3/2}^{-1}$ initial states. They are illustrated in Fig. \ref{fig:gamma}.
To the best of my knowledge, the Auger-Meitner decay widths of these systems have
not been presented in the literature so far.

Despite the difference in absolute numbers, the decay widths
of the different initial states show the same pattern. The decay width of
the $(n-1)p_{1/2,1/2}^{-1}$ initial state is lowest, while the decay widths
of the $(n-1)p_{3/2,1/2}^{-1}$ and
$(n-1)p_{3/2,3/2}^{-1}$ initial states are close and significantly higher than
for the $(n-1)p_{1/2,1/2}^{-1}$ initial state.
In case of the strontium atom, the decay width of the $p_{3/2}$ initial state
is approximately 12 times larger than the decay with of the $p_{1/2}$ initial state.
In radium, the corresponding factor is 380.
Furthermore, the decay width average of the $p_{3/2}$ initial state is increased 
by \unit[114]{\%} and \unit[236]{\%} for strontium and radium, respectively.
We can therefore observe an increase
of the decay width difference of the $p_{3/2}$ initial state both
compared to the $p_{1/2}$ and spinfree initial state for heavier atoms.
How can this result be understood?

\begin{figure}[h]
 \centering
 \includegraphics[width=\columnwidth]{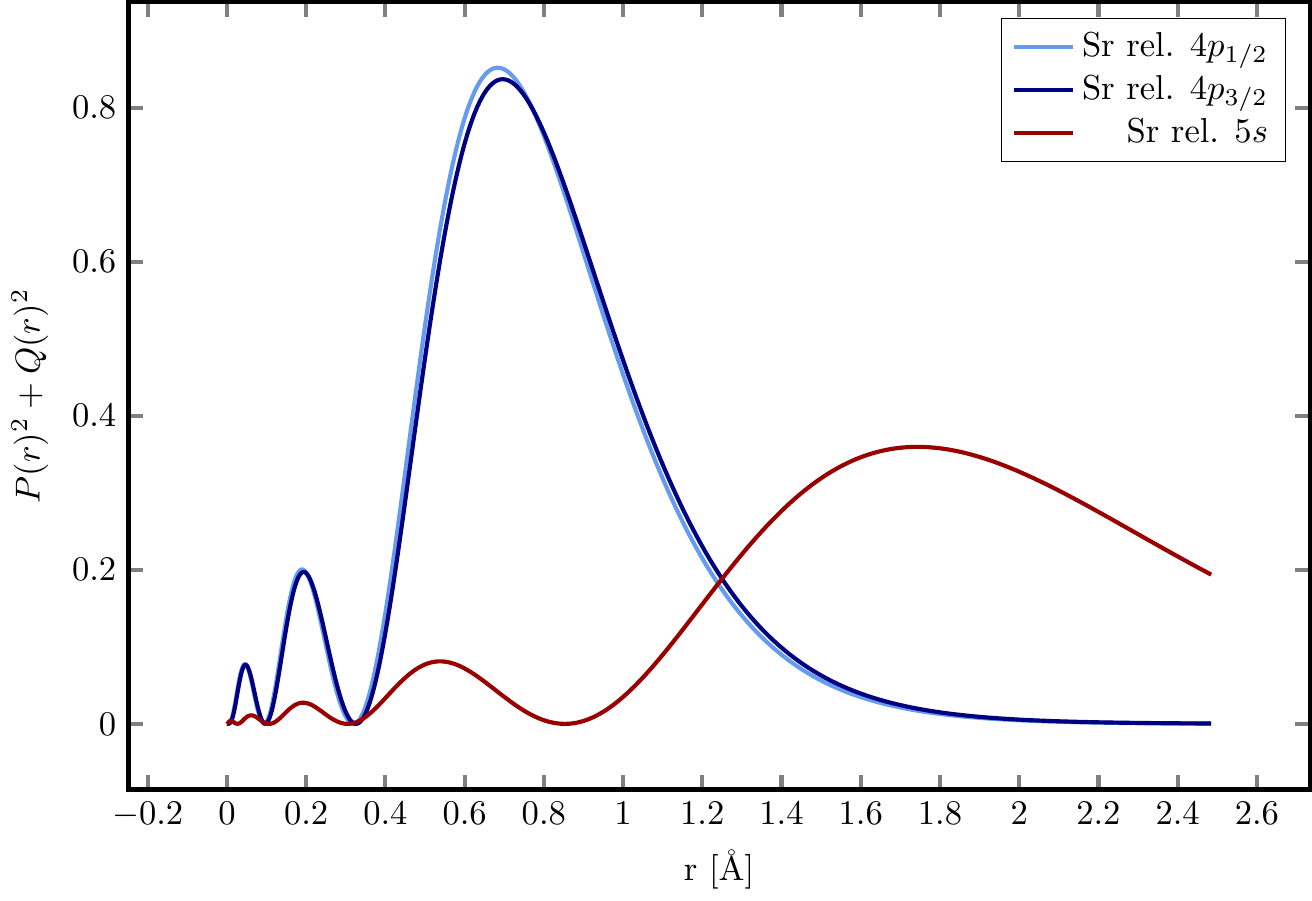}\\
 \includegraphics[width=\columnwidth]{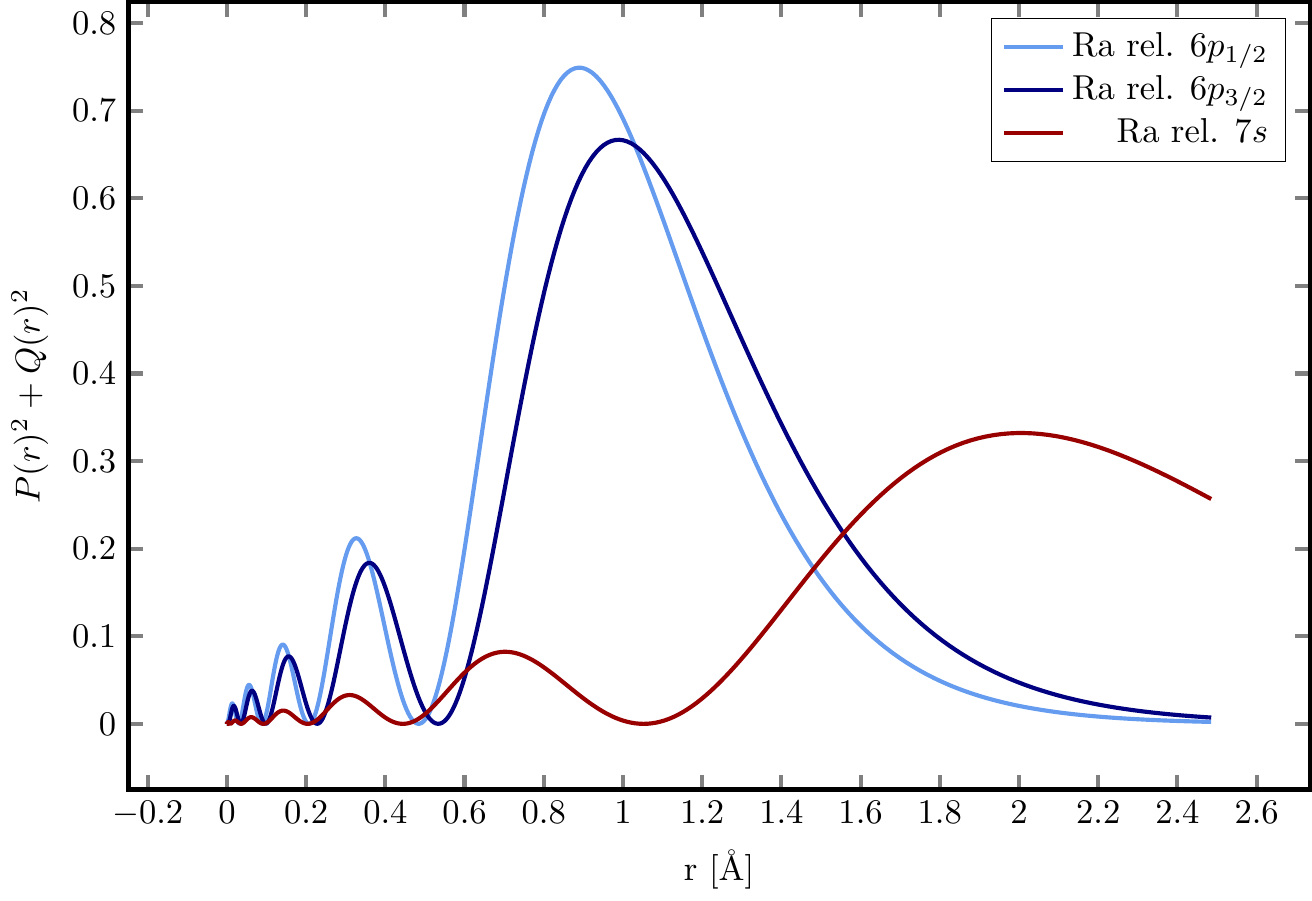}\\
 \caption{Radial densities of the orbitals of the $(n-1)p^5s^2$ ions
          involved in the Auger-Meitner decay.
          The expectation value of the electrons position of the $(n-1)p_{1/2}$
          orbital is lower than of the respective $(n-1)p_{3/2}$
          orbitals. The $ns$ orbitals of the ions experience a stronger
          contraction the those of the atom (not shown here).}
 \label{fig:radial_pure}
\end{figure}

Considering our previous findings about the role
of scalarrelativistic effects on electronic decay widths \cite{Fasshauer15_1},
we inspect the radial densities of the $(n-1)p$ and the $ns$ orbitals of the
strontium and radium ions,
which we assume to be involved in the decay process (see
Fig.~\ref{fig:radial_pure}).
In case of the strontium atom, the radial densities of the $4p_{1/2}$ and the
$4p_{3/2}$ orbitals are almost identical. But for the radium atom, the radial
density of the  $6p_{1/2}$ orbital is closer to the nucleus than
the radial density of the $6p_{3/2}$ orbital.
Because the Auger-Meitner decay rates crucially depend on the overlap of the
involved orbitals, the decay widths of an $(n-1)p_{3/2}^{-1}$ initial state
can be expected to be higher than the decay widths of an $(n-1)p_{1/2}^{-1}$
initial state. These findings are reflected in the decay widths shown in
Fig.~\ref{fig:gamma} and Table \ref{tab:widths} and are consistent with the
observations of the noble gas Auger-Meitner processes in Ref. \cite{Fasshauer15_1}.

\begin{table}[h]
 \centering
 \caption{Radial expectation values $\langle r \rangle$ of
          orbitals involved in the Auger-Meitner process
          for different configurations
          given in {\AA}ngstr{\"o}m.
}
 \begin{tabular}{llrrrrr}
  \toprule
   $\langle r \rangle$ &     config. &   $(n-1)p_{1/2}$ & $(n-1)p_{3/2}$ & $(n-1)d_{3/2}$ & $(n-1)d_{5/2}$ & $ns_{1/2}$\\
  \midrule
   \multirow{2}{*}{Sr} & $4p^55s^2$ & 0.794 &     0.809    &      --        &        --      &  1.981\\
       &      $4p^54d5s$ &            0.802 &     0.818    &    1.332       &      1.375     &  2.062\\
   \multirow{2}{*}{Ra} & $6p^57s^2$ & 0.996 &     1.114    &      --        &        --      &  2.244\\
       &      $6p^56d7s$ &            1.001 &     1.120    &    1.768       &      1.821     &  2.290\\
  \bottomrule
 \end{tabular}
 \label{tab:widths}
\end{table}

However, the simulations of Auger-Meitner processes in other earthalkaline elements
have shown that correlation effects are important for the correct
description of these elements' decay widths. Both the investigations of
the Auger-Meitner process following primary ionization
from the $2p$ orbitals of calcium \cite{Nikkinen05}
as well as from the $4d$ orbitals of barium \cite{Rose80}
showed the necessity to include excitations from the valence $s$ orbital to
the $d$ orbitals, which are unpopulated in the ground state, in the
description of the initial state.
In our case, this would require to include the following configurations in our
simulations:
 $(n-1)p^{-1} \,ns^2$,
 $(n-1)p^{-1} \,(n-1)d \, ns$ and
 $(n-1)p^{-1} \,(n-1)d^2$ .    
{Thereby, a fast Coster-Kronig decay is enabled equally for both the $l-1/2$ as
well as the $l+1/2$ case.}

Indeed, the analysis of the ADC eigenvectors of the initial states of both
strontium and radium showed
that beyond the single and main $1h$ contribution of the respective $p$ orbital, the
initial state is mainly characterized by $2h1p$ configurations of the
$(n-1)p^{-1} \,ns \, (n-1)d$ kind. They
are therefore automatically included in our simulations.
The
{relativistic}
FanoADC-Stieltjes
{implementation}
is, however, limited
to second order perturbations in the wavefunction
and therefore does not go beyond $2h1p$ configurations. This means that the
$(n-1)p^{-1} \,(n-1)d^2$ configurations are not taken into account in this work.

\begin{figure}[h]
 \centering
 \includegraphics[width=\columnwidth]{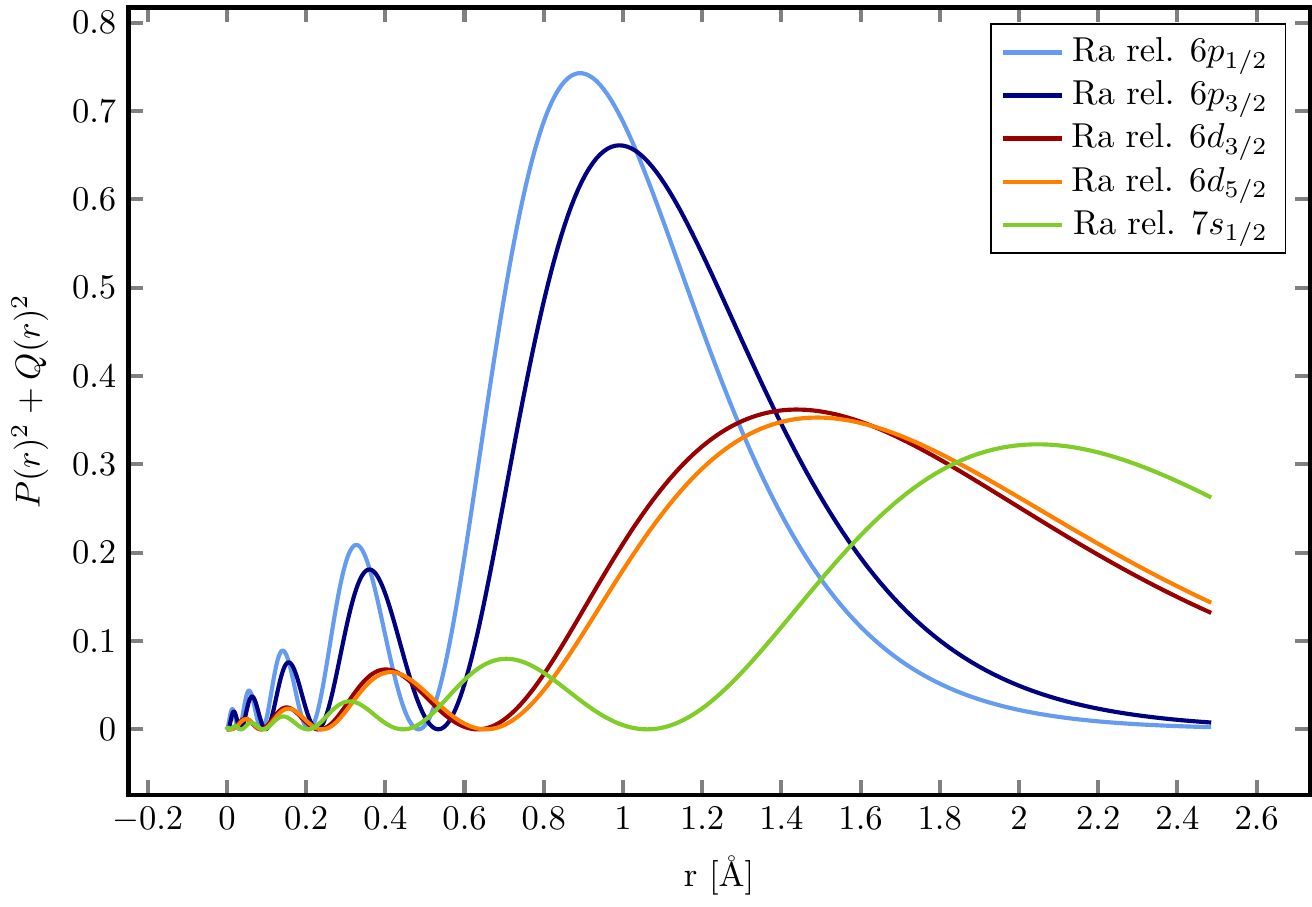}
 \caption{Radial densities of the radium orbitals of the ionic $6p^5 6d 7s$
          configuration involved in the Auger-Meitner decay. The overlap of the radial
          densities of the $6p$ orbitals with the radial densities of the $6d$ orbitals
          is more pronounced than with the radial density of the $7s$ orbital.}
 \label{fig:radial_ra}
\end{figure}

How do these configurations affect the decay widths? Since the overlap of the
orbitals involved in the decay determine the decay widths, we show the
radial densities of the orbitals involved in the decay of the $6p^5 6d 7s$
configuration of radium in Fig.~\ref{fig:radial_ra}.
The $6p$ orbitals are slightly more contracted than in the $6p^{-1} \,7s^2$
configuration, while the $7s$ orbital is slightly decontracted by a difference
of the electron's distance from the nucleus $\Delta \langle r \rangle$
of \unit[0.046]{\AA}.
However, the $6d$ orbitals, which are involved
in the Auger-Meitner process of this configuration, show a large overlap with both
the $6p$ and the $7s$ orbitals. The corresponding decay width should therefore
be larger than the decay width of the $6p^{-1} \,7s^2$ configuration.
The prevalence of the $(n-1)p_{3/2}$ decay width over the $(n-1)p_{1/2}$
also holds for this configuration.
If the $(n-1)p^{-1} \,(n-1)d^2$ configurations
have non-negligible contributions,
the presented decay widths of Table~\ref{tab:widths} would be
lower bounds to the results of a more accurate decay width calculation.

The observed differences in the decay widths for different initial states and
Hamiltonians can therefore not only be accounted for by relativistic
effects, but different contributions of the $(n-1)p^{-1} \,(n-1)d \, ns$
configurations may also affect the result.
In this case, the required measure of the $(n-1)p^{-1} \, ns^2$ configuration's
contribution to the ionized initial state is the pole-strength,
which are listed for the different initial states in Table \ref{tab:widths}.
The analysis of the eigenvectors has shown the other significant contributions
to be of the $(n-1)p^{-1} \,ns \, (n-1)d$ kind. We therefore assume that
they are the only other contributions and that their contribution is $1-\text{ps}$.
For strontium, the pole-strengths are similar but not identical for all initial
states and Hamiltonians. 
We can therefore predominantly attribute the decay width difference between the
$p_{3/2}$ and $p_{1/2}$ initial state to spin-orbit coupling rather than the
differences in the contributions of the $4p^54d5s$ configuration.
For radium, however, the pole-strengths of the initial states differ.
The $6p_{1/2}$ initial state has a pole-strength of 0.78, while both the
initial state of the scalarrelativistic Hamiltonian and the $6p_{3/2}$ initial
state have much lower pole-strengths of 0.49 and 0.50, respectively.
The $6p^{-1} \,7s^2$ configuration is therefore predominant for the $6p_{1/2}$
initial state, while it is not for the $6p_{3/2}$ initial state.
We can therefore not attribute the discrepancy of the decay widths of the
$6p_{1/2}$ and $6p_{3/2}$ initial states to spin-orbit splitting alone.
However, the initial state determined with the spinfree
Hamiltonian has a pole-strength comparable to the $6p_{3/2}$ initial states,
but the decay width of the $6p_{3/2}$ initial state is increased by \unit[236]{\%}
compared to the spinfree result.
Hence, we can explain the increase of the decay width from the spinfree to the
$6p_{3/2}$ initial state by spin-orbit coupling.

Based on these consistent findings, we can formulate the following rule of thumb for
the relative Auger-Meitner decay widths:\\
Two ionized initial states that stem from the same non-relativistic configuration and
are split by spin-orbit coupling will have different decay widths, where the decay width
of the $l-\frac12$ initial state will be significantly lower than the decay width of
the $l + \frac12$ initial state.\\
In contrast to the effect imposed by scalarrelativistic effects, this rule
of thumb is independent of the angular moment quantum number $l$, because the
$l-\frac12$ will always be contracted more strongly than the corresponding
$l+\frac12$ orbital and the final state orbitals will be further away from
the nucleus than the initial state orbitals.

{
\begin{table*}[htb]
 \caption{Auger-Meitner (AM) decay widths and Coster-Kronig (CK) widths
          of $L_2$ and $L_3$ shells in \unit{eV} extracted from
          Ref. \cite{Chen81}.}
 \begin{tabular}{ccrrr}
 \toprule
 Z & Element & $\Gamma_\text{AM}(L_2)$ & $\Gamma_\text{AM+CK}(L_2)$ & $\Gamma_\text{AM}(L_3)$\\
 \midrule
 25 & Mn & 0.390 &  --   & 0.337 \\
 30 & Zn & 0.664 &  --   & 0.689 \\
 36 & Kr & 1.101 & 1.219 & 1.167 \\
 40 & Zr & 1.373 & 1.578 & 1.465 \\
 45 & Rh & 1.705 & 2.029 & 1.860 \\
 47 & Ag & 1.832 & 2.192 & 2.028 \\
 50 & Sn & 2.016 & 2.454 & 2.256 \\
 52 & Te & 2.128 & 2.614 & 2.406 \\
 54 & Xe & 2.236 & 2.767 & 2.557 \\
 56 & Ba & 2.336 & 2.912 & 2.696 \\
 60 & Nd & 2.456 & 3.060 & 2.897 \\
 63 & Eu & 2.534 & 3.146 & 3.041 \\
 67 & Ho & 2.611 & 3.233 & 3.211 \\
 70 & Yb & 2.662 & 3.285 & 3.340 \\
 74 & W  & 2.753 & 3.423 & 3.543 \\
 80 & Hg & 2.877 & 3.600 & 3.878 \\
 90 & Th & 3.060 & 3.877 & 4.464 \\
 92 & U  & 3.082 & 4.272 & 4.571 \\
 96 & Cm & 3.126 & 5.386 & 4.805 \\
 98 & Cf & 3.146 & 5.360 & 4.934 \\
100 & Fm & 3.168 & 5.586 & 5.060 \\
 \bottomrule
 \end{tabular}
 \label{tab:L}
\end{table*}
The question is whether this rule of thumb is applicable to the Auger-Meitner
processes of other elements. For this, we refer to the Auger-Meitner and
Coster-Kronig decay widths of $L$ and $M$ initial states extracted from Refs.
\cite{Chen81,Chen83,Chen80} and presented in Tables \ref{tab:L} -- \ref{tab:M45}.
We note that this data is limited to comparably heavy elements as simulations
of Auger-Meitner decay widths based on relativistic wavefunctions are unavailable
in the literature for lighter elements.\\
\begin{table*}[h]
 \caption{Auger-Meitner (AM) decay widths and Coster-Kronig (CK) widths
          of $M_2$ and $M_3$ shells in \unit{eV} extracted from
          Ref. \cite{Chen83}.}
 \begin{tabular}{ccrrrr}
 \toprule
 Z & Element & $\Gamma_\text{AM}(M_2)$ & $\Gamma_\text{AM+CK}(M_2)$ & $\Gamma_\text{AM}(M_3)$ & $\Gamma_\text{AM+CK}(M_3)$ \\
 \midrule
 67 & Ho &  0.687 &  9.484 & 0.977 &  9.823\\
 70 & Yb &  0.277 & 10.259 & 1.125 &  9.949 \\
 74 & W  &  0.839 & 11.870 & 1.342 & 10.529 \\
 78 & Pt &  1.257 & 13.148 & 1.569 &  9.735 \\
 80 & Hg &  1.313 & 13.580 & 1.673 & 10.166 \\
 85 & At &  1.403 & 13.597 & 1.990 & 10.879 \\
 88 & Ra &  1.634 & 14.169 & 2.170 & 10.667 \\
 90 & Th &  1.650 & 14.404 & 2.264 &  9.973 \\
 92 & U  &  1.727 & 14.557 & 2.375 & 10.235 \\
 95 & Am &  1.836 & 15.130 & 2.629 & 10.850 \\
 \bottomrule
 \end{tabular}
 \label{tab:M}
\end{table*}
It is readily seen that the formulated rule of thumb holds for the Auger-Meitner
for all cases but one. The latter is the Auger-Meitner process of manganese after
primary ionization of a $2p$ orbital (see Table \ref{tab:L}).
However, the rule of thumb is not correct, when Coster-Kronig processes contribute
unequally to the decay of the $l+1/2$ and $l-1/2$ initial states.
The $L3$ and the $M5$ initial states are unable to undergo a Coster-Kronig decay
and they, as well as the $M3$ initial state have one fast decay channel than their
$l-1/2$ counterpart.
One may expect a clear trend for the relative decay widths with the nuclear charge
$Z$, but this hypothesis is not backed up by the data.
\begin{table*}[ht]
 \caption{Auger-Meitner (AM) decay widths and Coster-Kronig (CK) widths
          of $M_4$ and $M_5$ shells
          in \unit{eV} extracted from
          Ref. \cite{Chen80}.}
 \begin{tabular}{ccrrr}
 \toprule
 Z & Element & $\Gamma_\text{AM}(M_4)$ & $\Gamma_\text{AM+CK}(M_4)$ & $\Gamma_\text{AM}(M_5)$\\
 \midrule
 70 & Yb & 1.408 & 2.445 & 1.460 \\
 74 & W  & 1.758 & 1.831 & 1.830 \\
 80 & Hg & 2.259 & 2.381 & 2.452 \\
 83 & Bi & 2.512 & 2.670 & 2.753 \\
 88 & Ra & 2.923 & 3.126 & 3.253 \\
 92 & U  & 3.203 & 3.528 & 3.611 \\
 96 & Cm & 3.646 & 4.089 & 4.075 \\
100 & Fm & 4.000 & 4.488 & 4.583 \\
 \bottomrule
 \end{tabular}
 \label{tab:M45}
\end{table*}
}

{Based on the above reasoning, one would also expect}
this rule of thumb to extend to ICD and
Electron Transfer Mediated Decay (ETMD) processes as well. The ICD
decay widths partly depend on the transition dipole moment between the
initial and final state of the initially ionized unit, which in return increases
with the overlap of the electron densities of the involved orbitals.
The ETMD decay widths are determind by the orbital overlap of two different units,
which is larger for orbitals further away from the nucleus.
Hence the same argumentation as for the Auger-Meitner process
{should be possible}
in both cases.
{A validation will be left for future work.}


\section{Conclusions}
\label{section:conclusions}

We have presented lower bounds for
decay widths for the Auger-Meitner process initiated by
ionization from the $(n-1)p$ orbitals of strontium and radium.
Through analysis of results from different Hamiltonians and initial state
eigenvectors as well as radial densities of orbitals involved in the
Auger-Meitner process, we were able to show the importance of configuration interaction
in this specific case and the effect of spin-orbit coupling on decay widths
of electronic decay processes in general.
We condensed our findings into the following rule of thumb for the decay widths
of electronic decay processes:
Two ionized initial states that stem from the same non-relativistic configuration and
are split by spin-orbit coupling will have different decay widths, where the decay width
of the $l-\frac12$ initial state will be significantly lower than the decay width of
the $l + \frac12$ initial state.
{We have tested this rule of thumb against Auger-Meitner decay widths available in the
literature and could thereby validate it for the majority
of cases.}

\section{Acknowledgements}
The author would like to thank the audience of the REHE 2017 conference for
raising the main question addresed in this article
and acknowledges funding from the Villum foundation.

\clearpage


\end{document}